\documentclass[prb,aps,superscriptaddress,showpacs,reprint]{revtex4-1}
\usepackage{graphicx}
\usepackage{subfigure}
\usepackage{amsmath}
\usepackage{amssymb}
\usepackage{dsfont}

\usepackage{multirow}
\usepackage{rotating}

\usepackage{braket}
\usepackage{color} % use colored text in latex

\usepackage{hyperref}
\hypersetup{
colorlinks=true,final=true,
        linkcolor=blue,
        citecolor=blue,
        filecolor=blue,
        filecolor=blue,
        urlcolor=blue,
        pdfauthor={P. Kurpiers},
        pdftitle={Deterministic Generation of Remote Entanglement Using Microwave Photons}
}

\newcommand{\ha}{\hat{a}}

\newcommand{\had}{\hat{a}^\dagger}
\newcommand{\hb}{\hat{b}}
\newcommand{\hbd}{\hat{b}^\dag}

\newcommand{\hH}{\hat{H}}

\newcommand{\ketbra}[1]{\left| #1 \right>\left< #1 \right|}

\DeclareMathOperator{\sech}{sech}

\begin{document}
%%%%%%%%%%%%%%%%%%%%%%%%%%%%%%%%%%
%90 characters / 2 lines
%%%%%%%%%%%%%%%%%%%%%%%%%%%%%%%%%%
\title{Deterministic Quantum State Transfer and Generation of Remote Entanglement \\ using Microwave Photons}
\author{P.~Kurpiers}\altaffiliation{These authors contributed equally to this work.}\affiliation{Department of Physics, ETH Z\"urich, CH-8093 Z\"urich, Switzerland.}
\author{P.~Magnard} \altaffiliation{These authors contributed equally to this work.}\affiliation{Department of Physics, ETH Z\"urich, CH-8093 Z\"urich, Switzerland.}
\author{T.~Walter}\affiliation{Department of Physics, ETH Z\"urich, CH-8093 Z\"urich, Switzerland.}
\author{B.~Royer}\affiliation{Institut Quantique and D\'epartment de Physique, Universit\'e de Sherbrooke, Sherbrooke, Qu\'ebec J1K 2R1, Canada}
\author{M.~Pechal}\affiliation{Department of Physics, ETH Z\"urich, CH-8093 Z\"urich, Switzerland.}
\author{J.~Heinsoo}\affiliation{Department of Physics, ETH Z\"urich, CH-8093 Z\"urich, Switzerland.}
\author{Y.~Salath\'e}\affiliation{Department of Physics, ETH Z\"urich, CH-8093 Z\"urich, Switzerland.}
\author{A.~Akin}\affiliation{Department of Physics, ETH Z\"urich, CH-8093 Z\"urich, Switzerland.}
\author{S.~Storz}\affiliation{Department of Physics, ETH Z\"urich, CH-8093 Z\"urich, Switzerland.}
\author{J.-C.~Besse}\affiliation{Department of Physics, ETH Z\"urich, CH-8093 Z\"urich, Switzerland.}
\author{S.~Gasparinetti}\affiliation{Department of Physics, ETH Z\"urich, CH-8093 Z\"urich, Switzerland.}
\author{A.~Blais}\affiliation{Institut Quantique and D\'epartment de Physique, Universit\'e de Sherbrooke, Sherbrooke, Qu\'ebec J1K 2R1, Canada}\affiliation{Canadian Institute for Advanced Research, Toronto, Canada}
\author{A.~Wallraff}\affiliation{Department of Physics, ETH Z\"urich, CH-8093 Z\"urich, Switzerland.}
\date{\today}
%\begin{abstract}
%\end{abstract}

\maketitle

%%%%%%%%%%%%%%%%%%%%%%%%%%%%%%%%%%
% abstract: 200-300 words
%%%%%%%%%%%%%%%%%%%%%%%%%%%%%%%%%%
\textbf{
Sharing information coherently between nodes of a quantum network is at the foundation of distributed quantum information processing. In this scheme, the computation is divided into subroutines and performed on several smaller quantum registers connected by classical and quantum channels~\cite{Cirac1999}. A direct quantum channel, which connects nodes deterministically, rather than probabilistically, is advantageous for fault-tolerant quantum computation because it reduces the threshold requirements and can achieve larger entanglement rates~\cite{Jiang2007}. Here, we implement deterministic state transfer and entanglement protocols between two superconducting qubits~\cite{Wallraff2004} fabricated on separate chips. Superconducting circuits constitute a universal quantum node~\cite{Reiserer2015} capable of sending, receiving, storing, and processing quantum information~\cite{Eichler2012b,Johnson2010,Wenner2014,DiCarlo2009}. Our implementation is based on an all-microwave cavity-assisted Raman process~\cite{Pechal2014} which entangles or transfers the qubit state of a transmon-type artificial atom~\cite{Koch2007} to a  time-symmetric itinerant single photon. We transfer qubit states at a rate of $50 \, \rm{kHz}$ using the emitted  photons which are absorbed at the receiving node with a probability of $98.1 \pm 0.1 \%$ achieving a transfer process fidelity of $80.02 \pm 0.07 \%$. We also prepare on demand remote entanglement with a fidelity as high as $78.9 \pm 0.1 \%$. Our results are in excellent agreement with numerical simulations based on a master equation description of the system. This deterministic quantum protocol has the potential to be used as a backbone of surface code quantum error correction across different nodes of a cryogenic network to realize large-scale fault-tolerant quantum computation~\cite{Fowler2010,Horsman2012} in the circuit quantum electrodynamic (QED) architecture.}
%%%%%%%%%%%%%%%%%%%%%%%%%%%%%%%%%%%
% Introduction/ our implementation %1500 words main text
%%%%%%%%%%%%%%%%%%%%%%%%%%%%%%%%%%%

\begin{figure*}[t]
\centering
\includegraphics{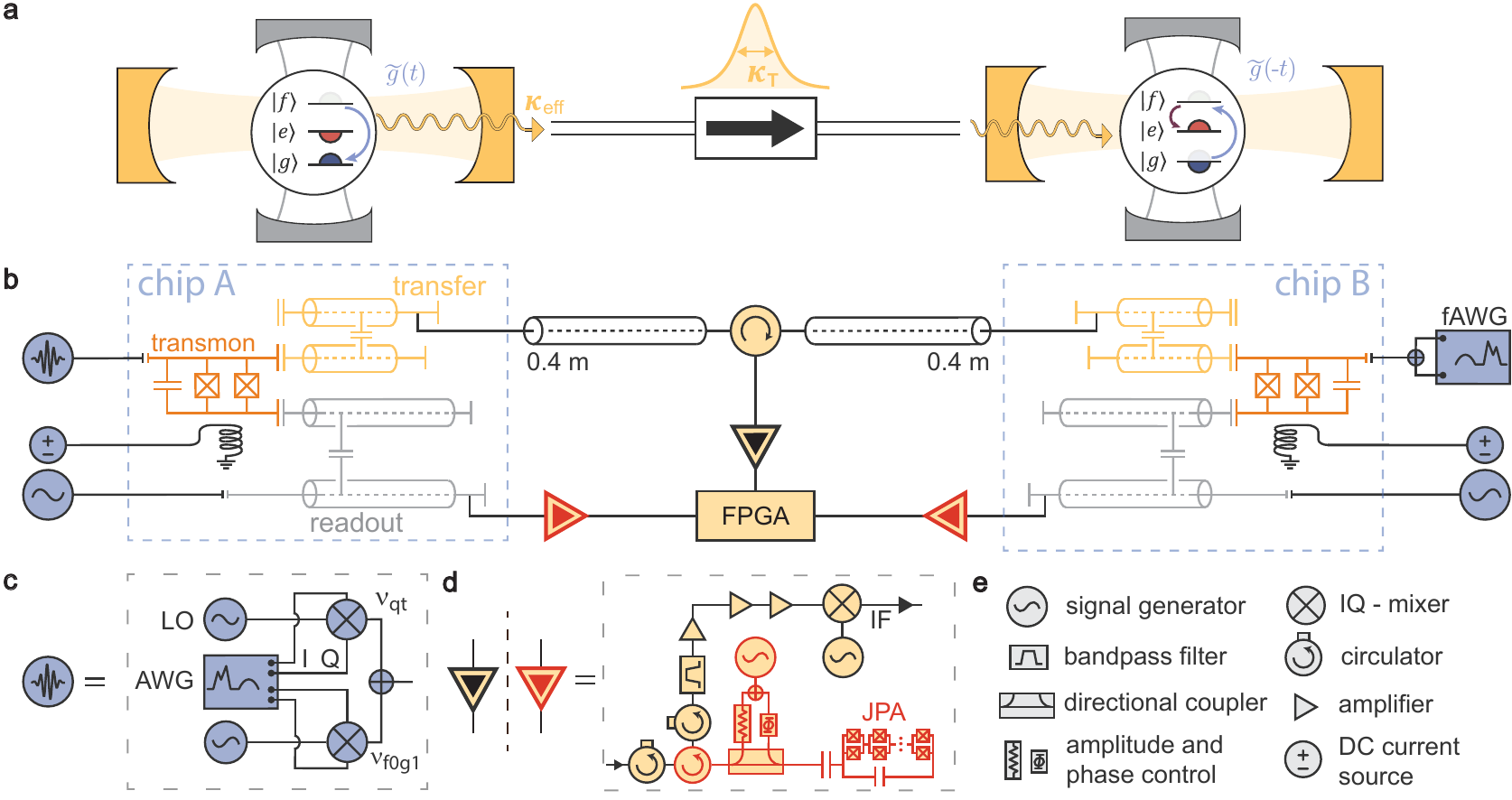}
\caption{\textbf{Schematic and measurement setup.} \textbf{a,} Quantum optical schematic of a deterministic unidirectional entanglement protocol between two cavity QED nodes of a quantum network. At the first node, a three level system is prepared in its second excited state $\ket{f}$ (grey half-circle) and coherently driven ($\tilde{g}(t)$, blue arrow) to $\ket{g}$  (blue half-circle) creating the transfer cavity field $\ket{1}$ (light yellow). The cavity field couples into the directional quantum channel with rate $\kappa_{\rm{T}}$ as a single photon wavepacket with an effective bandwidth $\kappa_{\rm{eff}}$ (yellow hyperbolic secant shape). In the second quantum node, the time reversed drive $\tilde{g}(-t)$ transfers the excitation from $\ket{g}$ to $\ket{f}$ in the presence of the transferred photon field $\ket{1}$. Finally, the protocol is completed with a transfer pulse between $\ket{f}$ and $\ket{e}$ (red half-circle) to return to the qubit subspace. Additionally, each three level system is coupled to a readout cavity (grey).  \textbf{b,} Circuit QED implementation of the system depicted in \textbf{a}. At each node, a transmon (orange) is coupled to two $\lambda/4$ coplanar waveguide resonator and Purcell filter circuits~\cite{Walter2017}, acting as the transfer (yellow) and readout (grey) cavities respectively. A directional quantum channel is realized using a semi-rigid coaxial cable and circulator connecting to the output port of the transfer circuit Purcell filter at each node. \textbf{c-e,} details of the circuit QED implementation. \textbf{c,} Combined qutrit ($\nu_{\rm{qt}}$) and $\ket{f,0}$ to $\ket{g,1}$ transition ($\nu_{\rm{f0g1}}$) microwave drive using single side-band modulation with in-phase (I) and quadrature (Q) mixers driven by a local oscillator (LO) and with an envelope defined by an arbitrary waveform generator (AWG) for node A. On node B these drives are directly synthesized by a fast AWG (fAWG) with $25 \; \rm{GS/s}$. \textbf{d,} Schematic of microwave detection lines (black). All detections lines consist of two isolators, a bandpass-filter, a cryogenic amplifier (HEMT) and two room-temperature amplifiers followed by a filter and analogue down-conversion to an intermediate frequency of 250~MHz. The down-converted signal is lowpass-filtered, digitized using an analogue-to-digital converter and recorded using a field-programmable gate array (FPGA). The readout lines include an additional Josephson parametric amplifier (JPA) circuit (red elements) between the first two isolators. The JPA is pumped by a signal generator and the reflected pump signal from the JPA is cancelled at a directional coupler using amplitude and phase ($\Phi$) controlled destructive interference. }
\label{fig:ProtocolSetup}
\end{figure*}

Remote entanglement has been realized probabilistically in heralded~\cite{Chou2007,Moehring2007,Hofmann2012,Bernien2013,Delteil2016} and unheralded protocols~\cite{Julsgaard2001,Matsukevich2006,Ritter2012,Roch2014} (see Appendix~\ref{app:litOverview} for details). A fully deterministic entanglement protocol  \cite{Cirac1997} utilizing a stationary atom coupled to a single mode cavity in remote quantum nodes  is more challenging to realize \cite{Ritter2012}. This protocol uses a coherent drive to entangle the state of an atom with the field of the cavity. The cavity is coupled to a directional quantum channel into which the field is emitted as a time-symmetric single photon. This photon travels to the receiving node where it is ideally absorbed with unit probability, using a time reversed coherent drive (Fig.~\ref{fig:ProtocolSetup}~a). In addition to establishing entanglement between the nodes, this direct transfer of quantum information naturally offers the possibility to transmit an arbitrary qubit state from one node to the other.

In our adaptation of this scheme (Fig.~\ref{fig:ProtocolSetup}~b) to the circuit QED architecture, each quantum node is composed of a superconducting transmon qubit with transition frequency $\nu_\mathrm{ge}^{\mathrm{A}} = 6.343\;\mathrm{GHz}$ ($\nu_\mathrm{ge}^{\mathrm{B}} = 6.093\;\mathrm{GHz}$) dispersively coupled to two coplanar microwave resonators, analogous to an atom in two cavities. One resonator is dedicated to dispersive qubit readout and the second one to excitation transfer. The transfer resonator of the two nodes have a matched frequency $\nu_{\rm{T}} = 8.400\;\mathrm{GHz}$ and a large bandwidth $\kappa_{\rm{T}}/2\pi \sim 11\;\mathrm{MHz}$ (see Appendix~\ref{app:Parameters}). All resonators are coupled to a dedicated filter, to protect the qubits from Purcell decay~\cite{Reed2010,Jeffrey2014,Walter2017}. An external coaxial line, bisected with a circulator, connects the transfer circuits of both nodes. With this setup, photons can be routed from node A to B, and from node B to a detection line. To generate a controllable light-matter interaction, we apply a coherent microwave tone to the transmon that induces an effective interaction $\tilde{g}(t)$ between states $\ket{f,0}$ and $\ket{g,1}$ with tunable amplitude and phase~\cite{Pechal2014,Zeytinoglu2015}. Here $\ket{s,n}$ denotes a Jaynes-Cummings dressed eigenstate with the transmon in state $\ket{s}$, where $\ket{g}$, $\ket{e}$ and $\ket{f}$ are its three lowest energy eigenstates, and $\ket{n}$ the Fock state of the transfer resonator. This interaction swaps an excitation from the transmon to the transfer resonator, which then couples to a mode propagating towards node B. By controlling $\tilde{g}(t)$ (see Appendix~\ref{app:Drive}), we shape the itinerant photon to have a time-reversal symmetric envelope $\phi(t)=\frac{1}{2}\sqrt{\kappa_{\rm{eff}}}\sech(\kappa_{\rm{eff}}t/2)$, with an adjustable photon bandwidth $\kappa_{\rm{eff}}$  limited only by $\kappa_{\rm{T}}$. By inducing the reverse process $\ket{f,0} \leftrightarrow \ket{g,1}$ with the time reversed amplitude and phase profile of $\tilde{g}(t)$ we absorb the itinerant photon with the transmon at node B. Ideally, this procedure returns all photonic modes to their vacuum state.

\begin{figure*}[t]
\centering
\includegraphics{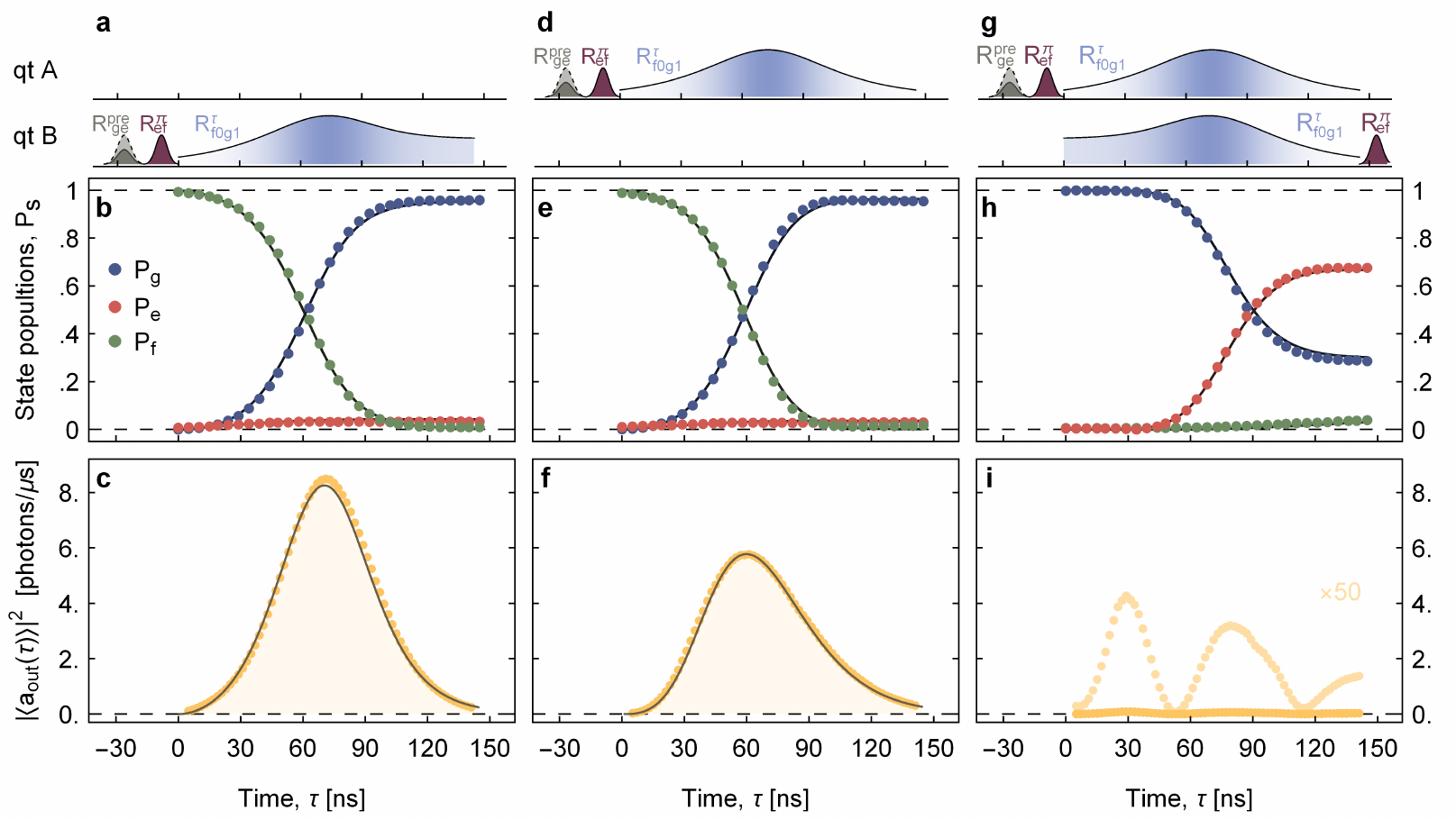}
\caption{\textbf{Emission, transfer and absorption of a single photon.} The transmon at node B (\textbf{a}) and node A (\textbf{d}) are prepared in the state $\ket{f}$ using Gaussian DRAG microwave pulses $\rm{R}^{\pi}_{\rm{ge}}$ and $\rm{R}^{\pi}_{\rm{ef}}$. We characterize (dots) the time dependence ($\tau$) of the qutrit populations $\rm{P}_{\rm{g,e,f}}$ (\textbf{b}, \textbf{e}) while driving the $\ket{f,0}$ to $\ket{g,1}$ transition (f0g1). The phase (white-blue shading) of the f0g1 drive is modulated to compensated the drive-induced quadratic ac Stark shift. The mean field amplitude squared $|\braket{\rm{{a}_{\rm{out}}(\tau)}}|^2$ of the travelling photons emitted from node B (\textbf{c}) and node A (\textbf{f}) is obtained for the emitted photon state $(\ket{0}+\ket{1})/\sqrt{2}$. The effective photon bandwidths are $\kappa^{\rm{A}}_{\rm{eff}}/2\pi=10.4 \, \rm{MHz}$ and $\kappa^{\rm{B}}_{\rm{eff}}/2\pi=10.6 \, \rm{MHz}$. The solid lines in \textbf{b, c, e, f, h, and i} are results of master equation simulations (see text for details). The time dependence of $\rm{P}_{\rm{s}}$ when executing the excitation transfer scheme (\textbf{g}) from qubit A to qubit B (\textbf{h}) are extracted simultaneously with the amplitude of the emitted field from node A. \textbf{i} shows the remaining $|\braket{\rm{{a}_{\rm{out}}(\tau)}}|^2$ (light yellow x50) during the absorption process.
}
\label{fig:Evolution}
\end{figure*}

%%%%%%%%%%%%%%%%%%%%%%%%%%%%%%%%%%%
% Results
%%%%%%%%%%%%%%%%%%%%%%%%%%%%%%%%%%%
To characterize the excitation transfer, we start by initializing the transmon in its ground state~\cite{Magnard2017} followed by a sequence of two $\pi$-pulses ($\rm{R}_{\rm{ge}}^{\pi}$, $\rm{R}_{\rm{ef}}^{\pi}$), used to prepare the transmon at node B in state $\ket{f,0}$. Next, we induce the effective coupling $\tilde{g}(t)$ with a modulated drive $\rm{R}_{\rm{f0g1}}^{\rm{\tau}}$ to emit a symmetric photon~\cite{Pechal2014} (Fig.~\ref{fig:Evolution}~a). We vary the instantaneous frequency of $\rm{R}_{\rm{f0g1}}^{\rm{\tau}}$, to compensate for the drive amplitude dependent ac-Stark shift of the $\ket{f,0} \leftrightarrow \ket{g,1}$ transition~\cite{Magnard2017} (see Appendix~\ref{app:Drive}). Here, and in all following measurements, the population of the transmon states are extracted using single-shot readout with a correction to account for measurement errors (see Appendix~\ref{app:Readout}). The population of the three lowest levels of the transmon $P_{\rm{g,e,f}}$ is measured immediately after truncating the emission pulse $\rm{R}_{\rm{f0g1}}^{\rm{\tau}}$ at time $\tau$ (see Fig.~\ref{fig:Evolution}~b). In this way, we observe that the transmon smoothly evolves from $\ket{f}$ to $\ket{g}$ during the emission process. The emitting transmon eventually reaches a ground state population $P_\mathrm{g} = 95\%$ which puts an upper bound to the emission efficiency.

To verify that the emitted photon envelope has the targeted shape and bandwidth $\kappa_\mathrm{eff}/2\pi = 10.4\; \mathrm{MHz}$, we repeat the emission protocol with an initial transmon state $ (\ket{g}+\ket{f})/\sqrt{2}$ and measure the averaged electric field amplitude $\langle a_\mathrm{out}(t)\rangle \propto \phi(t)$ of the emitted photon state $(\ket{0}+\ket{1})/\sqrt{2}$ using heterodyne detection~\cite{Bozyigit2011} (Fig.~\ref{fig:Evolution}~c). We prepare this photon state because of its non-zero average electric field~\cite{Pechal2014}. Repeating the emission protocol from node A, leads to similar dynamics of the transmon population (see Fig.~\ref{fig:Evolution}~e). The emitted photon state (Fig.~\ref{fig:Evolution}~f) has, however, a lower integrated power $\int |\langle a_\mathrm{out}(t)\rangle |^2 dt$ compared to emission from node B, due to a loss of $23.0 \pm0.5\%$ between the remote nodes (see Appendix~\ref{app:Loss}). The loss is extracted from the ratio of integrated photon powers for emission from nodes B and A.The photon emitted from node A changes shape when it reflects off node B due to the response function of its transfer resonator before being detected.

Finally, we measure the population of transmon B during the absorption of a single-photon emitted from A. We apply a $\pi$-pulse on transmon B right before the measurement to map $\ket{f}$ to $\ket{e}$. The excited state population, shown in Fig.~\ref{fig:Evolution}~h, smoothly rises before saturating at $P_\mathrm{e}^{\rm{sat}}$= 67.6 \%. This saturation level defines the total excitation transfer efficiency from node A to B which is reached here in only $180\;\mathrm{ns}$. From the ratio of the emitted photon integrated power in the absence (Fig.~\ref{fig:Evolution}~i) or presence (Fig.~\ref{fig:Evolution}~f) of the absorption pulse, the absorption efficiency is determined to be as high as $98\%$.

We perform master equation simulations (MES), shown as solid lines in Fig.~\ref{fig:Evolution}, of the excitation transfer experiments, using the time offset between the nodes as the only adjustable parameter (see Appendix~\ref{app:MES}). The excellent agreement between the MES and the data demonstrates a high level of control over the emission and absorption processes and an accurate understanding of the experimental imperfections. According to the MES these imperfections are accurately accounted for by decoherence and photon loss.

\begin{figure}[h]
\centering
\includegraphics{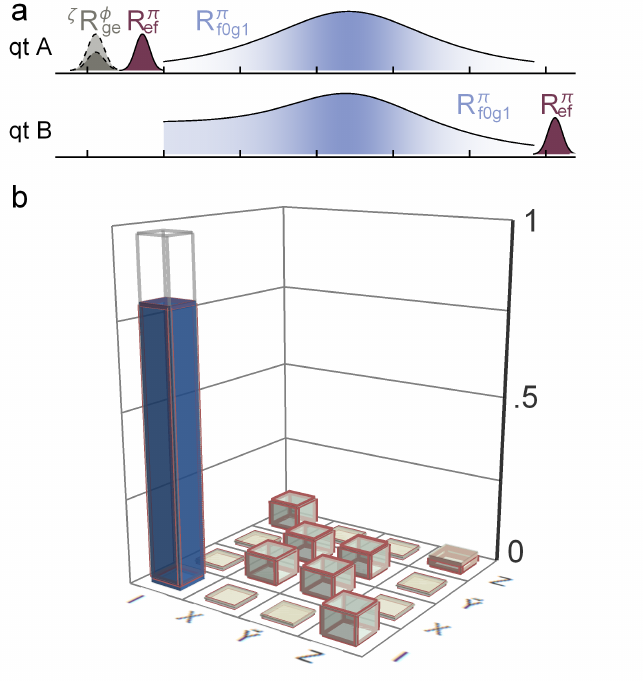}
\caption{\textbf{Quantum state transfer.} \textbf{a,} Pulse scheme used to characterize the qubit state transfer between the two nodes. We prepare six mutually unbiased input states with rotations $^{\rm{x}} \rm{R}^{0}_{\rm{ge}}$, $^{\rm{x}} \rm{R}^{\pi/2}_{\rm{ge}}$, $^{\rm{x}} \rm{R}^{-\pi/2}_{\rm{ge}}$, $^{\rm{y}} \rm{R}^{\pi/2}_{\rm{ge}}$, $^{\rm{y}} \rm{R}^{-\pi/2}_{\rm{ge}}$ and $^{\rm{x}} \rm{R}^{\pi}_{\rm{ge}}$ at node A (denoted by $^{\zeta} \rm{R}^{\phi}_{ge}$ where $\zeta$ is the rotation axis). \textbf{b, } We experimentally obtain (coloured bars) a process matrix with a fidelity of $\mathcal{F}_{\rm{p}}=80.02 \pm 0.07 \%$ relative to the ideal identity operation. The grey and red wire frames show the ideal and the master equation simulation of the absolute values of the process matrix, respectively. The trace distance between the measurement and the simulation is 0.014.}
\label{fig:stateTransfer}
\end{figure}

We demonstrate the use of the presented protocol to deterministically transfer an arbitrary qubit state from node A to node B. This is realized by preparing transmon B in state $\ket{g}$, applying a $\rm{R}^{\pi}_{\rm{ef}}$ on transmon A, followed by the emission/absorption pulse and finally a rotation $\rm{R}^{\pi}_{\rm{ef}}$ on transmon B. We characterize this quantum state transfer by reconstructing its process matrix $\chi$ with quantum process tomography (Fig.~\ref{fig:stateTransfer}~b). We prepare all six mutually unbiased qubit basis-states~\cite{Enk2007} at node A, transfer them to node B, and reconstruct the transferred state using quantum state tomography (QST) (see Appendix~\ref{app:QST}). The process fidelity is $\mathcal{F}_{\rm{p}}=\rm{Tr}(\chi\chi_{\rm{ideal}}) = 80.02 \pm 0.07 \%$, well above the limit of $1/2$ that could be achieved using local gates and classical communication only. The process matrix $\chi_{\rm{sim}}$ calculated with the MES, depicted with red wire frames in Fig.~\ref{fig:stateTransfer}~b, agrees well with the data, as suggested by the small trace distance $\sqrt{\rm{Tr}\left[(\chi-\chi_\mathrm{sim})^2\right]} = 0.014$.

\begin{figure}[t]
\centering
\includegraphics{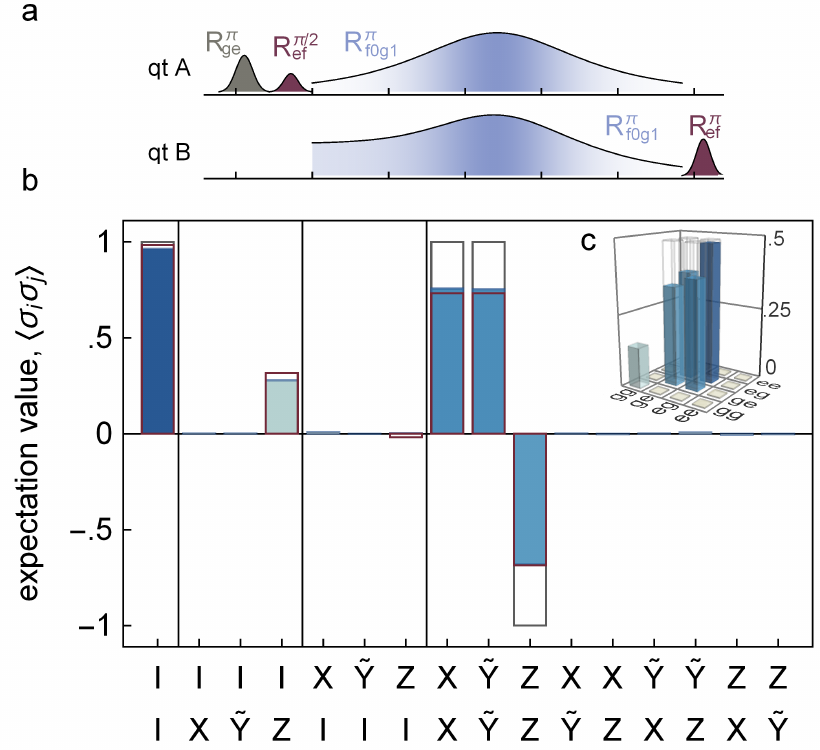}
\caption{\textbf{Remote entanglement generation.} \textbf{a,} Pulse scheme to generate deterministic remote entanglement between node A and B. \textbf{b, } Expectation values of two-qubit Pauli operators and \textbf{c, }reconstructed density matrix after execution of the remote entanglement protocol. \textbf{b,} The coloured bars indicate the measurement results, the ideal expectation values for the Bell state $\ket{\psi^+}=(\ket{ge}+\ket{eg})/\sqrt{2}$ are shown in grey wire frames and the results of a master equation simulation in red. We calculate a fidelity of $F=78.9 \pm 0.1 \%$ well explained by the photon loss and decoherence.}
\label{fig:entanglementGE}
\end{figure}

Finally, we use the excitation transfer to deterministically generate an entangled state between nodes A and B. The protocol starts by preparing transmon A and B in states $(\ket{e}+\ket{f})/\sqrt{2}$ and $\ket{g}$, respectively, and by applying the emission/absorption pulses followed by a rotation $\rm{R}^{\pi}_{\rm{ef}}$ on transmon B to generate the entangled Bell state $\ket{\psi^+}=(\ket{e_\mathrm{A},g_\mathrm{B}}+\ket{g_\mathrm{A},f_\mathrm{B}})/\sqrt{2}$. As leakage to the $\ket{f}$ level at both nodes leads to errors in the two-qubit density matrix reconstruction, we measure the full two-qutrit state $\rho_{3 \otimes 3}$ using QST (see Appendix~\ref{app:QST}).
For illustration purposes, we display the two-qubit density matrix $\rho_{m}$ (Fig.~\ref{fig:entanglementGE}~b and c), consisting of the two-qubit elements of $\rho_{3 \otimes 3}$.
We find a state fidelity compared to the ideal Bell state $\mathcal{F}^{s}_{\ket{\psi^+}}=\braket{\psi^+|\rho_{\rm{m}}|\psi^+}=78.9 \pm 0.1 \%$, and a concurrence $\mathcal{C}(\rho_{\rm{m}})=0.747 \pm 0.004$ (see Appendix~\ref{app:twoqutritEntanglement} for a detailed discussion). The state $\rho_\mathrm{sim}$ calculated from the MES of the entanglement protocol (red wireframe in Fig.~\ref{fig:entanglementGE}) results in a small trace distance $\sqrt{\rm{Tr}\left[(\rho_\mathrm{m}-\rho_\mathrm{sim})^2\right]}= 0.024$. The excellent agreement between the experimental and numerical results suggest that photon loss and finite coherence times of the transmons are the dominant sources of error, accounting for $12.5\%$ and $11\%$ infidelity, respectively.

Using transmons with relaxation and coherence times of $\rm{T}_{\rm{1ge}} = \rm{T}_{\rm{2ge}}= 30 \, \mu \rm{s}$, $\rm{T}_{\rm{1ef}}=\rm{T}_{\rm{2ef}} = 20 \, \mu \rm{s}$, and with an achievable $12\%$ loss  between the nodes, this protocol would allow deterministic generation of remote entangled states with fidelity $93\%$, at the threshold for surface code quantum error correction across different nodes~\cite{Fowler2010,Horsman2012,Perseguers2013,Campbell2017}. 
In addition, the protocol can be extended to generate deterministic heralded remote entanglement, utilizing the three-level structure of the transmons and encoding quantum information in different time bins to detect photon loss events, which would extend its functionality for quantum network applications~\cite{Reiserer2015}.
These perspectives indicate that the approach demonstrated here can serve as the basis for fault-tolerant quantum computation in the circuit QED architecture using distributed cryogenic nodes.

During writing of this manuscript we became aware of related work~\cite{Campagne-Ibarcq2017,Axline2017}.

\acknowledgments
This work is supported by the European Research Council (ERC) through the 'Superconducting Quantum Networks' (SuperQuNet) project, by the National Centre of Competence in Research 'Quantum Science and Technology' (NCCR QSIT), a research instrument of the Swiss National Science Foundation (SNSF), by ETH Zurich and NSERC, the Canada First Research Excellence Fund and the Vanier Canada Graduate Scholarships. 

\section*{Author contributions}
The experiment was designed and developed by P.K., T.W., P.M. and M.P. The samples were fabricated by J.-C.B., T.W. and S.G. The experiments were performed by P.K., P.M. and T.W. The data was analysed and interpreted by P.K., P.M., B.R., A.B. and A.W. The FPGA firmware and experiment automation was implemented by J.H., Y.S., A.A., S.S, P.M. and P.K. The master equation simulation were performed by B.R., M.P., P.M. and P.K. The manuscript was written by P.K., P.M., T.W., B.R. and A.W. All authors commented on the manuscript. The project was led by A.W.

%%%%%%%%%%%%%%%%%%%%%%%%%%%%%%%%%%%%%%%%%%%%%%%%%%%%%%%%%%%%%%%%%%%%%%%%%%%%%%%%%%%%%%%%%%%%%%%%%%%%%%%
\appendix

\section{Literature Overview}
\label{app:litOverview}
\begin{figure}
\centering
\includegraphics{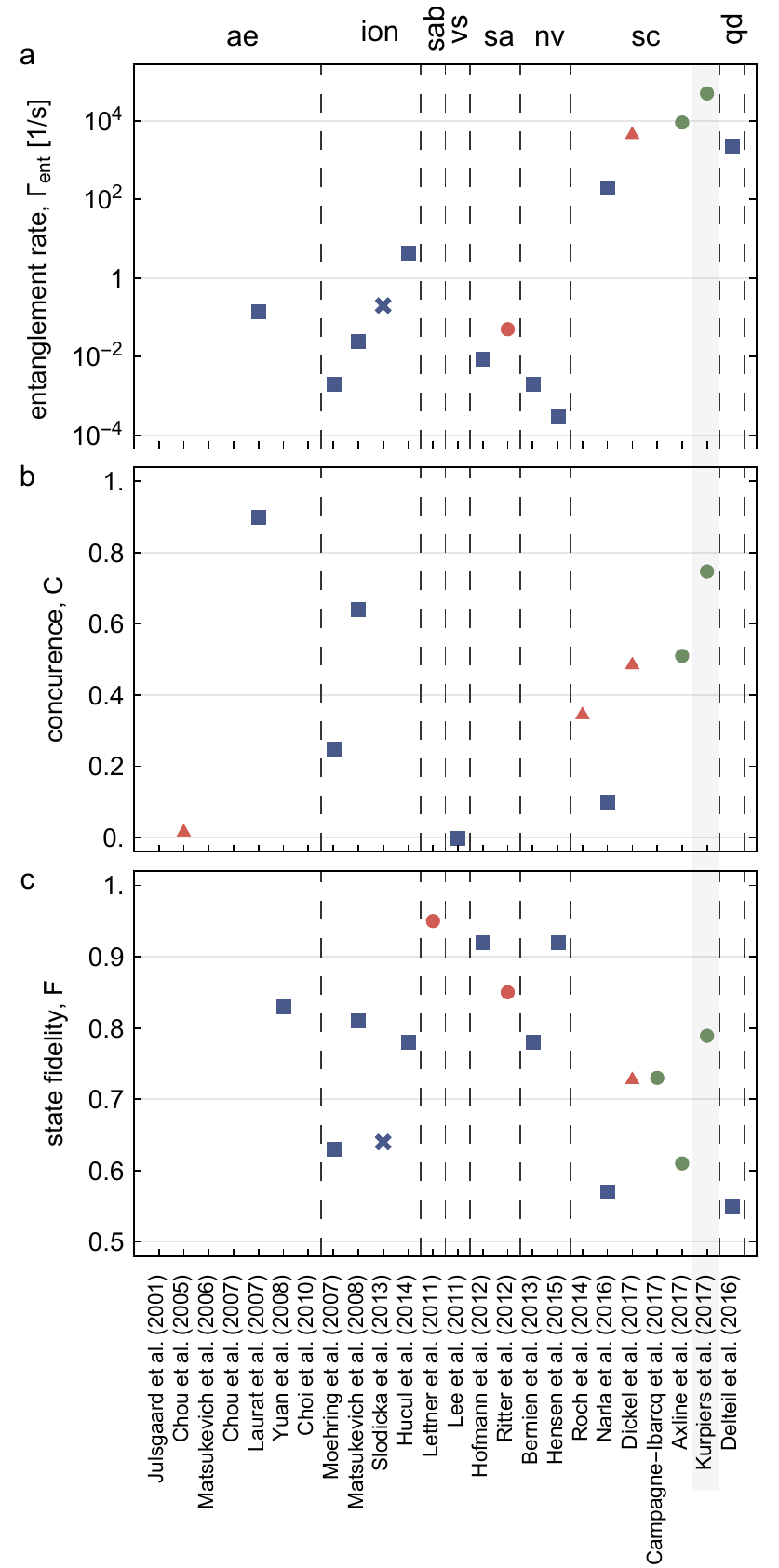}
\caption{\textbf{Overview of remote entanglement experiments.} \textbf{a,} entanglement rates  \textbf{b,} concurrence, \textbf{c,} entangled state fidelity. The experiments are sorted by physical system: atomic ensembles (ae)~\cite{Julsgaard2001,Chou2005,Matsukevich2006,Chou2007,Laurat2007,Yuan2008,Choi2010}, trapped ions (ion)~\cite{Moehring2007,Matsukevich2008,Slodicka2013,Hucul2014}, single atom~-~Bose Einstein condensate (sab)~\cite{Lettner2011},vibrational state of diamonds (vs)~\cite{Lee2011}, single atoms (sa)~\cite{Hofmann2012,Ritter2012}, nitrogen-vacancy (nv)  center~\cite{Bernien2013,Hensen2015}, superconducting circuits (sc)~\cite{Roch2014,Narla2016,Dickel2017,Campagne-Ibarcq2017,Axline2017} and quantum dots (qd)~\cite{Delteil2016}. The colours indicate probabilist unheralded (red), probabilist heralded (blue), deterministic unheralded (green) implementations. The plot markers indicate different schemes to realize the remote interaction: measurement induced (triangle), interference of two single photons on beam splitter (squares), single photon emission and detection (cross), direct transfer (pentagon), direct transfer with shaped photons (circles).}
\label{fig:LitOverview}
\end{figure}
We provide an overview of remote entanglement experiments performed in a range of physical systems using several different schemes listed in the caption of Fig.~\ref{fig:LitOverview}.

\section{Sample Parameters}
\label{app:Parameters}
\begin{table}[b]
\begin{tabular}{|c|rr|}
\hline 
 & \hfill{}Node A\hfill{} & \hfill{}Node B\hfill{}\tabularnewline
\hline 
$\nu_{\mathrm{R}}$ & 4.787~GHz & 4.780~GHz\tabularnewline
$\nu_{\mathrm{Rpf}}$ & 4.778~GHz & 4.780~GHz\tabularnewline
$\kappa_{R}/2\pi$ & 12.6~MHz & 27.1~MHz\tabularnewline
$\chi_{R}/2\pi$ & 5.8~MHz & 11.6~MHz\tabularnewline
$\nu_{\mathrm{T}}$ & 8.4005~GHz & 8.4003~GHz\tabularnewline
$\nu_{\mathrm{Tpf}}$ & 8.426~GHz & 8.415~GHz\tabularnewline
$\kappa_{T}/2\pi$ & 10.4~MHz & 13.5~MHz\tabularnewline
$\chi_{T}/2\pi$ & 6.3~MHz & 4.7~MHz\tabularnewline
$\nu_{\mathrm{ge}}$ & 6.343~GHz & 6.096~GHz\tabularnewline
$\alpha$ & -265~MHz & -308~MHz\tabularnewline
$T_{\mathrm{1ge}}$ & 4.9~$\mathrm{\mu s}$ & 4.6~$\mathrm{\mu s}$\tabularnewline
$T_{\mathrm{1ef}}$ & 1.6~$\mathrm{\mu s}$ & 1.4~$\mathrm{\mu s}$\tabularnewline
$T_{\mathrm{2ge}}$ & 3.4~$\mathrm{\mu s}$ & 2.6~$\mathrm{\mu s}$\tabularnewline
$T_{\mathrm{2ef}}$ & 2.1~$\mathrm{\mu s}$ & 0.9~$\mathrm{\mu s}$\tabularnewline
\hline 
\end{tabular}
\caption{\label{tab:ParameterSummary} Summary of device parameters for node A and B. With $\ell=\rm{R}, \, \rm{T}$, $\nu_{\ell}$ is the frequency of the coupling resonator, $\nu_{\mathrm{\ell pf}}$ the frequency of the Purcell Filter, $\kappa_{\ell}/2\pi$ the effective decay rate of the coupled resonator to the external feed line and $\chi_{\ell}/2\pi$ the dispersive coupling strength of the transmon readout or transfer circuit, respectively. }
\end{table}

The devices are identical to the one found in Ref.~\onlinecite{Walter2017} with only minor parameter modifications.  The $\lambda/4$ coplanar waveguide resonators and additional feed-lines are created from etched niobium on a sapphire substrate using standard photolithography techniques. We then define the transmon pads and junctions with electron-beam lithography and shadow evaporated aluminium with lift-off. We extract the parameters of the readout circuit (grey Fig.~\ref{fig:ProtocolSetup}~b) and transfer circuit (yellow Fig.~\ref{fig:ProtocolSetup}~b), as well as the coupling strength of the transmon to these circuits, with fits to the transmission spectra of the respective Purcell filter when the transmon is prepared in its ground and excited state using the technique and model as discussed in Ref.~\onlinecite{Walter2017}. Furthermore, the anharmonicity, the energy relaxation times and the coherence times of the qutrits are found using Ramsey-type measurements. Finally, we used miniature superconducting coils to thread flux through the SQUID of each transmon to tune their frequencies such that their transfer circuit resonator had identical frequencies. All relevant device parameters are summarized in Table~\ref{tab:ParameterSummary}.

%%%%%%%%%%%%%%%%%%%%%%%%%%%%%%%%%%%%%%%%%%%%%%%%%%%%%%%%%%%%%%%%%%%%%%%%%%%%%%%%%%%%%%%%%%%%%%%%%%%%%%%
\section{Microwave Drive Schemes}
\label{app:Drive}
\begin{figure}[b]
\centering
\includegraphics{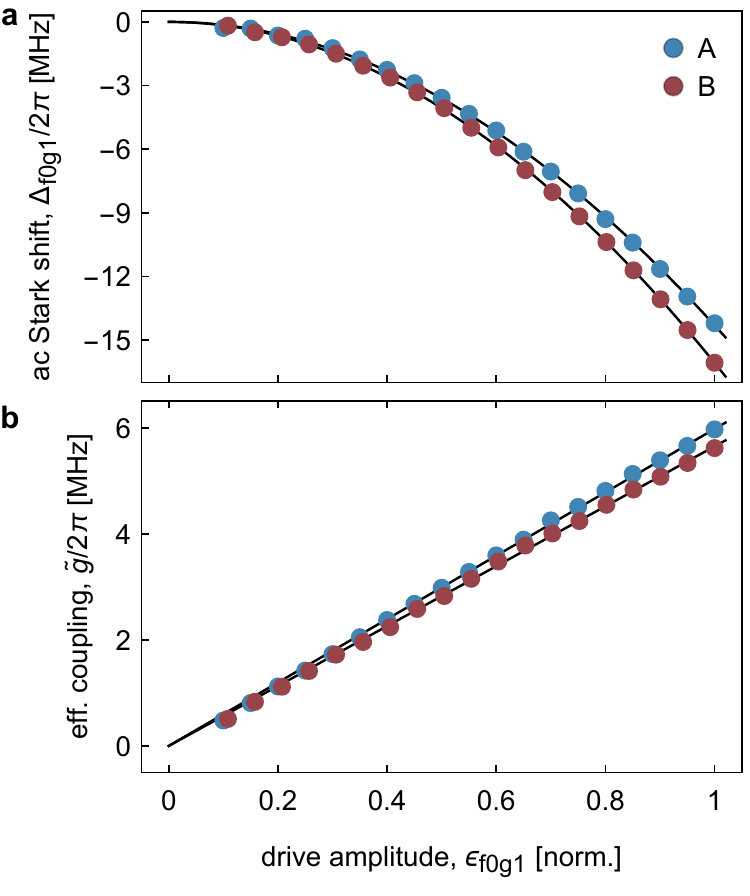}
\caption{\textbf{AC stark shift and Rabi rate of the $\ket{f,0}$ to $\ket{g,1}$ transition.}  Measurement (dots) of the ac Stark shift $\Delta_{\rm{f0g1}}/2\pi$ (\textbf{a}) and effective coupling $\tilde{g}/2\pi$ (\textbf{b}) of the $\ket{f0}$ to $\ket{g1}$ transition versus drive amplitude $\rm{\epsilon_{\rm{f0g1}}}$ for sample A and B. The solid lines in \textbf{a} (\textbf{b}) are quadratic (linear) fits to the data.}
\label{fig:calibF0G1}
\end{figure}
We use resonant Gaussian-shaped DRAG~\cite{Motzoi2009,Gambetta2011} microwave pulses of length $19.8 \, \rm{ns}$ and $16.8 \, \rm{ns}$ for $\rm{R}^{\pi}_{\rm{ge}}$ and $\rm{R}^{\pi}_{\rm{ef}}$ in order to swap populations between the $\ket{g}$ and $\ket{e}$ state and the $\ket{e}$ and $\ket{f}$ state respectively. We extract an averaged Clifford-gate fidelity for the $\ket{g}$ and $\ket{e}$ pulses of more than $99.2 \%$ for both transmon qubits, from randomized benchmarking experiments~\cite{Chow2009}.

We induce the effective coupling $\tilde{g}$ between states $\ket{f,0}$ and $\ket{g,1}$ by applying a microwave tone on the transmon with drive amplitude $\rm{\epsilon}$, at the resonance frequency of the transition $\nu_{\rm{f0g1}}^{\rm{\rm{A}}}=4.0219 \, \rm{GHz}$  and $\nu_{\rm{f0g1}}^{\rm{\rm{B}}}=3.4845 \, \rm{GHz}$. Following the procedure described in Refs.~\onlinecite{Magnard2017} and~\onlinecite{Pechal2014}, we calibrate the ac Stark shift of the transmon levels induced by the $\ket{f,0} \leftrightarrow\ket{g,1}$ drive, and extract the linear relation between the drive amplitude $\rm{\epsilon}$ and the effective coupling $\tilde{g}$ (see Fig.~\ref{fig:calibF0G1}). We adjust the phase of $\rm{\epsilon}$ based on the measured ac Stark shift in order to remain resonant with the driven transition. We calibrate our transmon drive lines to reach a maximum effective coupling $\rm{\tilde{g}}^{\rm{A}}/2\pi=6.0 \, \rm{MHz}$ and  $\rm{\tilde{g}}^{\rm{B}}/2\pi= 6.7 \, \rm{MHz}$ (Fig.~\ref{fig:calibF0G1} b).

\begin{figure*}[t]
\includegraphics{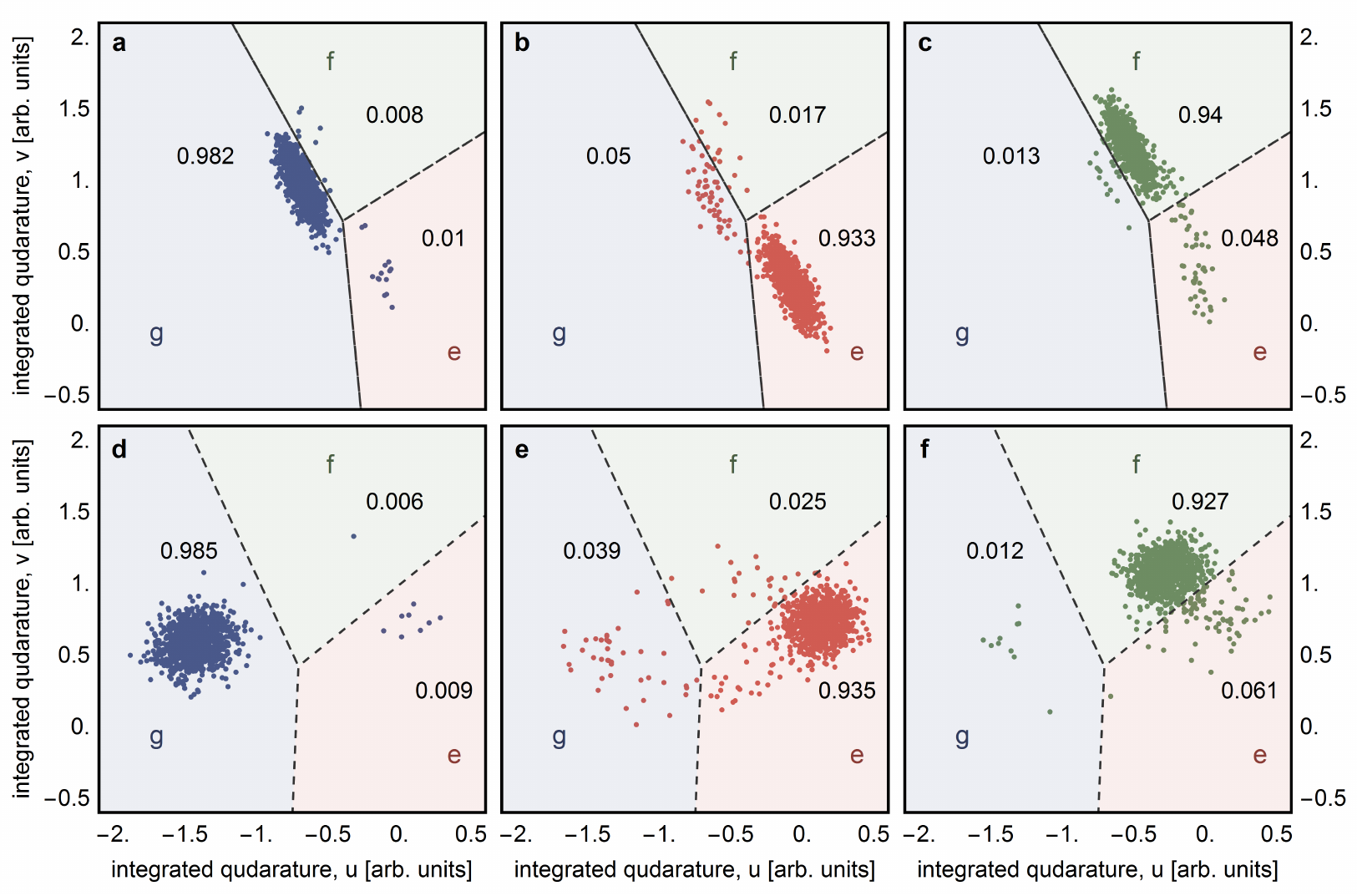}
\caption{\textbf{Qutrit single-shot readout characterization.} Scatter plot of the measured integrated quadrature values u, v for qutrit A (\textbf{a}-\textbf{c}) and B (\textbf{d}-\textbf{f}) when prepared in state $\ket{g}$ (blue), $\ket{e}$ (red), $\ket{f}$ (green), respectively. We plot only the first 1000 of the total 25000 repetitions for each state preparation experiment. The dashed lines are the qutrit state discrimination thresholds used to obtain the assignment probabilities (indicated numbers and also listed in Table~\ref{tab:QutritProbMatrix}).}
\label{fig:singleShot}
\end{figure*}

We generate photons with temporal shape $\phi(t)=\frac{1}{2}\sqrt{\kappa_{\rm{eff}}}\sech(\kappa_{\rm{eff}}t/2)$ by resonantly driving the $\ket{f,0} \leftrightarrow\ket{g,1}$ transition with
\begin{equation}
\rm{\tilde{g}}(t)= \frac{\kappa_{\rm{eff}}}{4 \cosh \frac{\kappa_{\rm{eff}}t}{2}} \frac{1-\rm{e}^{\kappa_{\rm{eff}}t}+(1+\rm{e}^{\kappa_{\rm{eff}}t})\kappa_{\rm{T}}/\kappa_{\rm{eff}}}{\sqrt{(1+\rm{e}^{\kappa_{\rm{eff}}t})\kappa_{\rm{T}}/\kappa_{\rm{eff}}-\rm{e}^{\kappa_{\rm{eff}}t}}}.
\label{eqn:driveshape}
\end{equation}
where $\kappa_{\rm{T}}$ is the coupling of the transfer resonator to the coaxial line, and $\kappa_{\rm{eff}}$ is determined by the strength and duration of the transfer pulse, and is constrained by $\kappa_{\rm{eff}}\leq\kappa_{\rm{T}}$. The dynamics are well described by a two-level model with loss, captured by the non-Hermitian Hamiltonian
	\begin{equation}
	H = 
	  \begin{bmatrix}
    0            & \tilde{g} \\
		\tilde{g}^*	 & i \kappa/2
		\end{bmatrix}
	\label{eq:f0g1Rabi}
	\end{equation}
which acts on states $\ket{f,0}$ and $\ket{g,1}$, analysed in a rotating frame. The non-Hermitian term $i \kappa/2$ accounts for photon emission, which brings the system to the dark state $\ket{g,0}$. One can show that using the effective coupling of Equation~(\ref{eqn:driveshape}) in the Hamiltonian~(\ref{eq:f0g1Rabi}) leads to the emission of a single photon with the desired temporal shape.

%%%%%%%%%%%%%%%%%%%%%%%%%%%%%%%%%%%%%%%%%%%%%%%%%%%%%%%%%%%%%%%%%%%%%%%%%%%%%%%%%%%%%%%%%%%%%%%%%%%%%%%
\section{Three-Level Single-Shot Readout}
\label{app:Readout}
The state of transmon A (B) is read out with a gated microwave tone, with frequency $\nu_\mathrm{d}^{\mathrm{A}} = 4.778\;\mathrm{GHz}$  ($\nu_\mathrm{d}^{\mathrm{B}} = 4.765\;\mathrm{GHz}$), applied to the input port of the Purcell filter. As depicted in Fig.~\ref{fig:ProtocolSetup}~b, the output signal is routed through a set of two circulators and a combiner and then amplified at $10\;\mathrm{mK}$ with $22\;(19.3)\;\mathrm{dB}$ gain using a Josephson parametric amplifier (JPA). The JPA pump tone is $2\;\rm{MHz}$ detuned from the measurement signal and has a bandwidth of $18.3\;(32)\;\mathrm{MHz}$. Using these JPAs we find a phase-preserving detection efficiency of $\eta_{\rm{2Q}} = 0.61 \; (0.60)$ for the full detection line.The signal is then further amplified by a high electron mobility transistor (HEMT) at $4\;\rm{K}$ and two low-noise amplifiers at room temperature. Next, the signal is analogue down-converted to $250\;\rm{MHz}$, lowpass-filtered, digitized by an analog-to-digital converter and processed by a field-programmable gate array (FPGA). Within the FPGA, the data is digitally down-converted to DC and the corresponding I and Q quadratures values are recorded during a window of 256~ns in 8~ns time steps. The FPGA trigger is timed so that the measurement window starts with the rising edge of the measurement tone. We refer to a recording of the I and Q quadrature of a measurement tone as a readout trace, $S(t)$.

We prepare the transmon in state  \(\ket{g}\), \(\ket{e}\) and \(\ket{f}\), 25000 times each and record the single-shot traces. Each trace is then integrated in post-processing, with two weight functions $w_1(t)$ and $w_2(t)$, to obtain the integrated quadratures $u = \int S(t) w_1(t) dt$ and $v = \int S(t) w_2(t) dt$. The collected and integrated traces form three Gaussian shaped clusters in the $u$-$v$ plane (Fig.~\ref{fig:singleShot}), that correspond to the Gaussian probability distributions of the trace when the qutrit is prepared in one of the three eigenstates. We model the probability distribution $(u,v)$ as a mixture of three Gaussian distributions, with density\begin{equation}
f(\vec{x})=\sum_s{\frac{A_\mathrm{s}}{2\pi \sqrt{|\Sigma|}} e^{-\frac{1}{2}(\vec{x}-\mu_\mathrm{s})^\top\cdot\Sigma^{-1}\cdot(\vec{x}-\mu_\mathrm{s})}}
\label{eqn:densityfunction}
\end{equation}
and estimate the parameters $A_\mathrm{s}$, $\mu_\mathrm{s}$ and $\Sigma$ with likelihood maximization. Based on these parameters, we divide the $u$-$v$ plane into three regions used to assign the result of the readout of the qutrit state (Fig.~\ref{fig:singleShot}). If an integrated trace is in the region labelled $s'$, we assign it state $s'$.
By counting the number of traces prepared in state $|s\rangle$ and assigned the value $s'$, we estimate the assignment probabilities $R_\mathrm{ss'}=P(s'|\,|s\rangle)$ (see Fig.~\ref{fig:singleShot}). We optimize the measurement power and signal integration time in order to minimize the measurement error probability $\frac{1}{6}||I-R||_1$. The optimum occurs with the measurement time $t_\mathrm{m}=112\;\mathrm{ns}$ and input power $P_\mathrm{in} = -24 \;\mathrm{dBm}$ for qutrit A and $t_\mathrm{m}= 216\;\mathrm{ns}$,   $P_\mathrm{in}=-25 \;\mathrm{dBm}$ for qutrit B. The total assignment error probability is approximately 5\% for both qutrits as seen in the assignment probability matrix compiled in Table~\ref{tab:QutritProbMatrix}.
\begin{table}[b]
\begin{centering}
\begin{tabular}{|c|ccc|c|ccc|c|}
\cline{1-4} \cline{6-9} 
\multicolumn{1}{|c}{} & \multicolumn{3}{c|}{Qutrit A} &  & \multicolumn{3}{c}{Qutrit B} & \tabularnewline
\cline{1-4} \cline{6-9} 
\multicolumn{1}{|c}{} & \hfill{}$\left|\mathrm{g}\right\rangle $\hfill{} & \hfill{}$\left|\mathrm{e}\right\rangle $\hfill{} & \hfill{}$\left|\mathrm{f}\right\rangle $\hfill{} &  & \hfill{}$\left|\mathrm{g}\right\rangle $\hfill{} & \hfill{}$\left|\mathrm{e}\right\rangle $\hfill{} & \multicolumn{1}{c}{\hfill{}$\left|\mathrm{f}\right\rangle $\hfill{}} & \tabularnewline
\cline{1-4} \cline{6-9} 
g & 98.2 & 5.0 & 1.3 &  & 98.5 & 3.9 & 1.2 & g\tabularnewline
e & 1.0 & 93.3 & 4.8 &  & 0.9 & 93.5 & 6.1 & e\tabularnewline
f & 0.8 & 1.7 & 94.0 &  & 0.6 & 2.5 & 92.7 & f\tabularnewline
\cline{1-4} \cline{6-9} 
\end{tabular}
\par\end{centering}
\caption{\label{tab:QutritProbMatrix} Probabilities of identifying prepared input states (columns) as the indicated output states (rows) for qutrit A and B. The diagonal elements show correct identification, the off-diagonal elements misidentifications.}
\end{table}
The probability $M_\mathrm{s'}$ to assign value $s'$ to a single shot measurement of a qutrit in state $\rho$ is given by 
\begin{equation}
M_\mathrm{s'} = P(s'|\rho) = \sum_s P(s'||s\rangle)\cdot \rho_\mathrm{ss}
\label{eq:assignProb}
\end{equation}
which can be expressed as $M=R\cdot\vec{\rho}_\mathrm{diag}$ where $\vec{\rho}_\mathrm{diag}$ is the vector consisting of the diagonal elements of $\rho$. The assignment probabilities $M$ are typically estimated from assignment counts and a first approach to estimate $\vec{\rho}_\mathrm{diag}$ is to equate $\vec{\rho}_\mathrm{diag}=M$. This approach is sensitive to measurement errors, but insensitive to state preparation errors. Setting $\vec{\rho}_\mathrm{diag}=R^{-1}\cdot M$ effectively accounts for the effect of single-shot readout error. However, this approach relies on the ability to estimate $R$ precisely and thus is sensitive to state-preparation error. With transmon reset infidelities of approximately $0.2\%$~\cite{Magnard2017}, and single qubit gate errors of $0.6\%$ (measured with randomized benchmarking), state preparation errors are expected to be lower than readout errors. For this reason, we chose to use the latter approach.

We note that the assignment probability matrix $R_\mathrm{s_\mathrm{A} s_\mathrm{B},s'_\mathrm{A} s'_\mathrm{B}}=P(s'_\mathrm{A}s'_\mathrm{B}|\,|s_\mathrm{A}s_\mathrm{B}\rangle) = P(s'_\mathrm{A}|\,|s_\mathrm{A}\rangle)\cdot P(s'_\mathrm{B}|\,|s_\mathrm{B}\rangle)$ can be obtained as the outer product of the single-qutrit assignment probability matrices (compiled in Table~\ref{tab:2QutritProbMatrix}) and that we can extend this formalism to correct for single-shot readout errors and extract the state populations of a two-qutrit system.

\begin{table}
\begin{centering}
\begin{tabular}{|c|ccccccccc|}
\hline 
 & $\left|\mbox{\ensuremath{\mathrm{gg}}}\right\rangle $ & $\left|\mbox{\ensuremath{\mathrm{ge}}}\right\rangle $ & $\left|\mbox{\ensuremath{\mathrm{gf}}}\right\rangle $ & $\left|\mbox{\ensuremath{\mathrm{eg}}}\right\rangle $ & $\left|\mbox{\ensuremath{\mathrm{ee}}}\right\rangle $ & $\left|\mbox{\ensuremath{\mathrm{ef}}}\right\rangle $ & $\left|\mbox{\ensuremath{\mathrm{fg}}}\right\rangle $ & $\left|\mbox{\ensuremath{\mathrm{fe}}}\right\rangle $ & $\left|\mbox{\ensuremath{\mathrm{ff}}}\right\rangle $\tabularnewline
\hline 
gg & 96.8 & 3.9 & 1.1 & 4.9 & 0.2 & 0.1 & 1.2 & 0.0 & 0.0\tabularnewline
ge & 0.9 & 91.9 & 6.0 & 0.0 & 4.7 & 0.3 & 0.0 & 1.2 & 0.1\tabularnewline
gf & 0.6 & 2.5 & 91.1 & 0.0 & 0.1 & 4.6 & 0.0 & 0.0 & 1.2\tabularnewline
eg & 1.0 & 0.0 & 0.0 & 91.9 & 3.7 & 1.1 & 4.7 & 0.2 & 0.1\tabularnewline
ee & 0.0 & 0.9 & 0.1 & 0.8 & 87.3 & 5.7 & 0.0 & 4.5 & 0.3\tabularnewline
ef & 0.0 & 0.0 & 0.9 & 0.6 & 2.4 & 86.5 & 0.0 & 0.1 & 4.4\tabularnewline
fg & 0.8 & 0.0 & 0.0 & 1.6 & 0.1 & 0.0 & 92.5 & 3.7 & 1.1\tabularnewline
fe & 0.0 & 0.7 & 0.0 & 0.0 & 1.6 & 0.1 & 0.8 & 87.9 & 5.8\tabularnewline
ff & 0.0 & 0.0 & 0.7 & 0.0 & 0.0 & 1.6 & 0.6 & 2.4 & 87.1\tabularnewline
\hline 
\end{tabular}
\par\end{centering}
\label{tab:2QutritProbMatrix}
\caption{Probabilities of identifying prepared input states (columns) as the indicated output states (rows) for all possible tuples of  two-qutrit basis state $\ket{g}$, $\ket{e}$ and $\ket{f}$. The diagonal elements show correct identification, the off-diagonal elements misidentifications.}
\end{table}

%%%%%%%%%%%%%%%%%%%%%%%%%%%%%%%%%%%%%%%%%%%%%%%%%%%%%%%%%%%%%%%%%%%%%%%%%%%%%%%%%%%%%%%%%%%%%%%%%%%%%%%
\section{Loss Estimation}
\label{app:Loss}
The loss on the printed circuit boards including  connectors is measured to be $2.5 \pm 1 \%$, of the coaxial cables of length $0.4 \, \rm{m}$ (each $4.0 \pm 0.1 \%$)~\cite{Kurpiers2017} and information provided by the manufacturer for the microwave circulator ($13 \pm 2 \%$).

%%%%%%%%%%%%%%%%%%%%%%%%%%%%%%%%%%%%%%%%%%%%%%%%%%%%%%%%%%%%%%%%%%%%%%%%%%%%%%%%%%%%%%%%%%%%%%%%%%%%%%%
\section{Master Equation Simulation}
\label{app:MES}
We model the transmons as anharmonic oscillators with annihilation (creation) operators $\hb_i$ ($\hbd_i$)~\cite{Koch2007}, where the subscript $i=A,B$ denotes the emitter and receiver samples, respectively. The transfer resonator annihilation (creation) operators are denoted $\ha_i$ ($\had_i$). Setting $\hbar = 1$, the driven Jaynes-Cummings Hamiltonian for sample $i$ is given by
\begin{equation}\label{eq:HamTransverse}
\begin{aligned}
\hH^i =&\; \omega_T^i \had_i \ha_i + \omega_{eg}^i \hbd_i \hb_i + \Omega^i(t)(\hb_i + \hbd_i)\\
& + g_T^i(\had_i \hb_i + \ha_i \hbd_i) - \frac{E_C^i}{2} \hbd_i \hbd_i \hb_i \hb_i,
\end{aligned}
\end{equation}
where $g_T^i$ denotes the coupling between the transmon and the transfer resonator, $E_C^i$ the charging energy of the transmon and $\Omega^i(t) =  \Omega^i \cos[\omega_d^i t + \varphi^i(t)]$ is the amplitude of the microwave drive inducing the desired coupling $\tilde g(t)$. Since the readout resonators do not play a role in the photon transfer dynamics, they are omitted from the Hamiltonian and the static Lamb shifts they induce are implicitly included in the parameters.

In order to make the effective coupling $\tilde g(t)$ between the $\ket{f,0}$ and $\ket{g,1}$ states apparent and to simplify the simulations, we perform a series of unitary transformations on Equation~(\ref{eq:HamTransverse}). First moving to a frame rotating at the drive frequency $\omega_d^i$, we then perform a displacement transformation $\hb_i \rightarrow \hb_i - \beta^i, \ha_i \rightarrow \ha_i - \gamma^i$ and choose $\beta^i,\gamma^i$ such that the amplitude of the linear drive terms is set to zero. Next, we perform a Bogoliubov transformation $\hb_i \rightarrow \cos(\Lambda^i)\hb_i - \sin(\Lambda^i)\ha_i, \ha_i \rightarrow \cos(\Lambda^i)\ha_i + \sin(\Lambda^i)\hb_i$, where $\tan(2\Lambda^i) = -2g_T^i/(\omega_{T}^i - \omega_{eg}^i + 2 E_C^i |\beta^i|)$ and, neglecting small off-resonant terms, obtain the resulting effective Hamiltonian
\begin{equation}\label{eq:HamTransfo}
\begin{aligned}
\hH^i_{\tilde g} =&\; \Delta_T^i \had_i \ha_i + \Delta_{eg}^i \hbd_i \hb_i + \frac{\alpha^i}{2}\hbd_i \hbd_i \hb_i \hb_i + \frac{K^i}{2} \had_i \had_i \ha_i \ha_i\\
& + 2\chi_T^i \had_i \ha_i \hbd_i \hb_i + \frac{1}{\sqrt 2}(\tilde g \hbd_i \hbd_i \ha_i + \tilde g^* \had_i \hb_i \hb_i),\\
\end{aligned}
\end{equation}
where $\alpha^i = -E_C^i \cos^4\Lambda$ is the transmon anharmonicity, $K^i = -E_C^i \sin^4\Lambda^i$ is the qubit-induced resonator anharmonicity, $\chi_T^i = - E_C^i \cos^2\Lambda^i\sin^2\Lambda^i $ is the dispersive shift, $\Delta_T^i = \omega_T^i \cos^2\Lambda^i  + (\omega_{ge}^i - 2 E_C^i |\beta^i|^2)\sin^2\Lambda^i - g_T^i\sin2\Lambda^i - \omega_d^i$ is the resonator-drive detuning and $\Delta_{eg}^i = \omega_{ge}^i - 2 E_C^i |\beta^i|^2) \cos^2\Lambda^i  + \omega_T^i \sin^2\Lambda^i + g_T^i\sin2\Lambda^i  - \omega_d^i$ is the qubit-drive detuning.
In Equation~(\ref{eq:HamTransfo}), the desired effective coupling $\tilde g^i = - E_C^i \beta^i \sqrt 2 \cos^2\Lambda^i \sin\Lambda^i$ between the $\ket{f,0}$ and $\ket{g,1}$ states is now made explicit.

\begin{figure*}[t]
\centering
\includegraphics{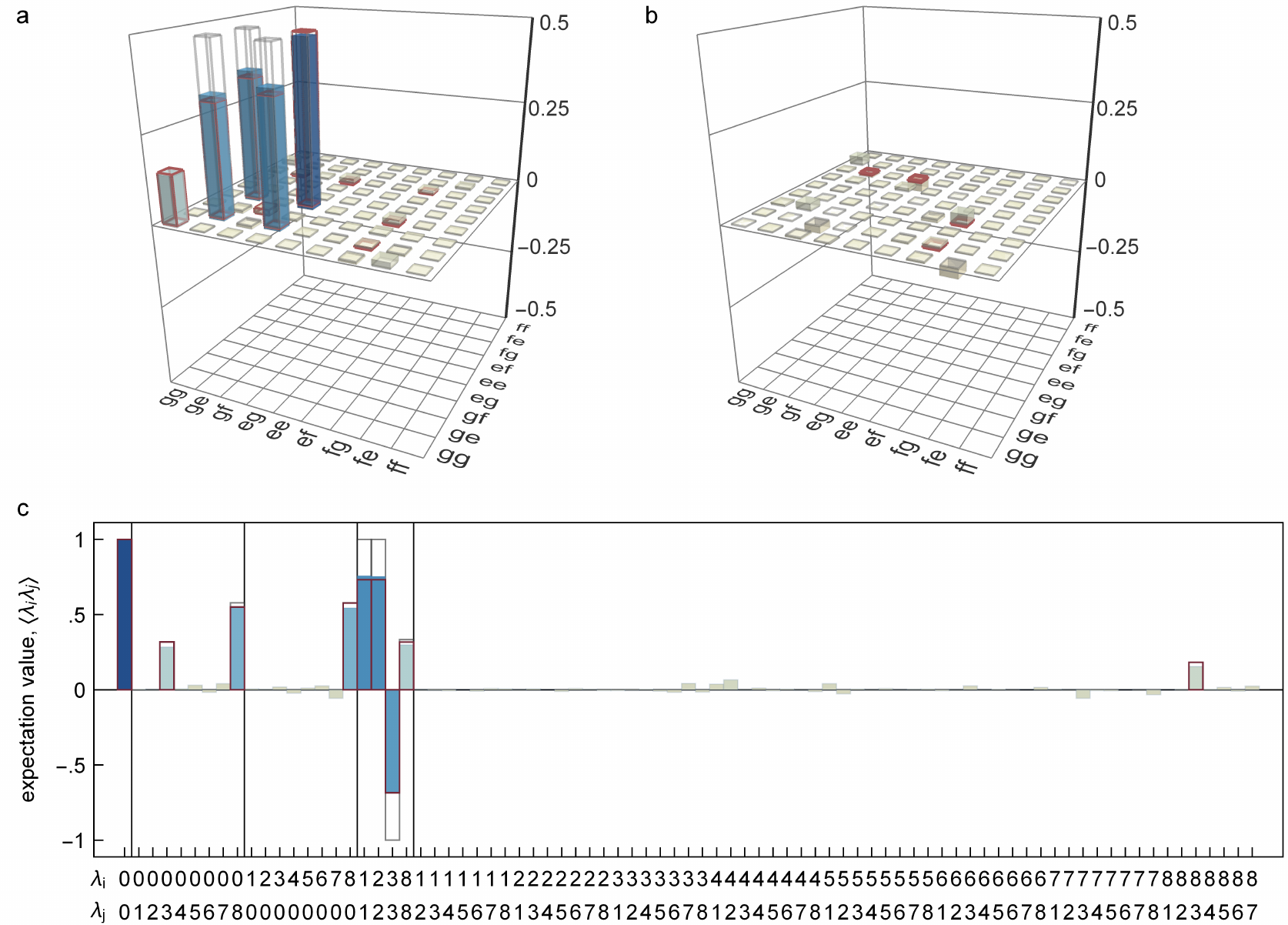}
\caption{\textbf{Characterization of a remotely entangled state.} We prepare a qubit-qubit entangled state between the distant quantum systems using the protocol described in the main text and perform two-qutrit state tomography: (\textbf{a}) real and (\textbf{b}) imaginary  part of the density matrix and (\textbf{c}) expectation values of the Gell-Mann operators $\lambda_k$. The ideal Bell state $\ket{\psi^+}$ is depicted with grey wire frames. The numerical master equation simulation is depicted in red wire frames. $\lambda_0$ denotes the identity operation, $\lambda_{1,2,3}$ the Pauli matrices $\sigma_{\rm{x}, \rm{y}, \rm{z}}^{ge}$ in the qubit (ge) subspace, $\lambda_{4,5}$ correspond to $\sigma_{\rm{x}, \rm{y}, \rm{z}}^{gf}$, $\lambda_{6,7}$ to  $\sigma_{\rm{x}, \rm{y}, \rm{z}}^{ef}$ and $\lambda_8$ is the diagonal matrix $(\sigma_{\rm{z}}^{ge}+2\sigma_{\rm{z}}^{ef})/\sqrt{3}$. The trace distance between the measurement and the simulation is 0.027.}
\label{fig:entanglementGEF}
\end{figure*}

Finally, moving to a frame rotating at $\Delta_T^i$ for the resonator and $\Delta_{eg}^i + \alpha^i/2$ for the transmon qubits, the combined effective Hamiltonian of the two samples can be written as
\begin{equation}\label{eq:Heff}
\begin{aligned}
\hH_{eff} =& \sum_{i= A,B} \left\{ - \frac{\alpha^i}{2} \hbd_i \hb_i +\frac{\alpha^i}{2} \hbd_i \hbd_i\hb_i\hb_i \right.\\
 &+ \frac{K^i}{2} \had_i \had_i \ha_i \ha_i + 2\chi_T^i \had_i \ha_i \hbd_i \hb_i\\
 &\left. + \frac{1}{\sqrt{2}}\left[\tilde g^i(t) \hbd_i \hbd_i \ha_i + \tilde g^i(t)^* \had_i \hb_i \hb_i\right]\right\}\\
 &-i\frac{\sqrt{\kappa_{T}^A \kappa_{T}^B \eta_c}}{2}(\ha_A \had_B - \had_A \ha_B),
\end{aligned}
\end{equation}
where $\eta_c$ is the photon loss probability of the circulator between the two samples.
Using this effective Hamiltonian, numerical results are obtained by integrating the master equation 
\begin{equation}\label{eq:ME}
\begin{aligned}
\dot \rho =& -i[\hH_{eff},\rho]\\
& + \kappa_T^A (1-\eta_c) \mathcal D[\ha_A]\rho + \mathcal D[\sqrt{\kappa_T^A \eta_c} \ha_A + \sqrt{ \kappa_T^B} \ha_B]\rho\\
& + \sum_{i=A,B} \left\{\kappa^i_{int}\mathcal D[\ha_i]\rho + \gamma_{1ge}^i\mathcal D\left[\ket{g}\bra{e}_i\right]\rho\right.\\
&\left.\;\;\;\;\;\; + \gamma_{1ef}^i\mathcal D\left[\ket{e}\bra{f}_i\right]\rho\right\}\\
& + \sum_{i=A,B} \left\{ \gamma_{\phi ge}^i\mathcal D\left[\ketbra{e}_i - \ketbra{g}_i\right]\rho \right.\\
&\left.\;\;\;\;\;\; + \gamma_{\phi ef}^i\mathcal D\left[\ketbra{f}_i - \ketbra{e}_i\right]\rho\right\},
\end{aligned}
\end{equation}
where $\mathcal D[\hat O]\bullet = \hat O\bullet \hat O^\dag  - \{\hat O^\dag \hat O,\bullet\}/2$ denotes the dissipation super-operator, $\kappa^i_{int}$ the internal decay rates of the resonators, $\gamma_{1nm}^i=1/T^i_{1nm}$ the decay rates of the transmon qubits between the $\ket{n}_i,\ket{m}_i$ states and $\gamma^i_{\phi nm}=1/2T_{1nm}^i - 1/T_{2nm}^i$ the dephasing rates between the $\ket{n}_i,\ket{m}_i$ states of the transmon qubits.
The last term in $\hH_{eff}$ combined with the resonator dissipators in the second line of the master equation~(\ref{eq:ME}), assure that the output of the emitter A is cascaded to the input of the receiver B~\cite{Gardiner1993,Carmichael1993} through a circulator with photon loss $\eta_c$.

%%%%%%%%%%%%%%%%%%%%%%%%%%%%%%%%%%%%%%%%%%%%%%%%%%%%%%%%%%%%%%%%%%%%%%%%%%%%%%%%%%%%%%%%%%%%%%%%%%%%%%%
\section{Quantum State and Process Tomography}
\label{app:QST}
Quantum state tomography of a single qutrit is performed by measuring the qutrit state population with the single-shot readout method described in Appendix~\ref{app:Readout}, after applying the following tomography gates: 
$^{\rm{x}} \rm{R}^{0}_{\rm{ge}}$, 
$^{\rm{x}} \rm{R}^{\pi/2}_{\rm{ge}}$, 
$^{\rm{y}} \rm{R}^{\pi/2}_{\rm{ge}}$, 
$^{\rm{x}} \rm{R}^{\pi}_{\rm{ge}}$, 
$^{\rm{x}} \rm{R}^{\pi/2}_{\rm{ef}}$, 
$^{\rm{y}} \rm{R}^{\pi/2}_{\rm{ef}}$, 
$(^{\rm{x}} \rm{R}^{\pi}_{\rm{ge}}\,^{\rm{x}} \rm{R}^{\pi/2}_{\rm{ef}})$, 
$(^{\rm{x}} \rm{R}^{\pi}_{\rm{ge}}\,^{\rm{y}} \rm{R}^{\pi/2}_{\rm{ef}})$ and 
$(^{\rm{x}} \rm{R}^{\pi}_{\rm{ge}}\,^{\rm{x}} \rm{R}^{\pi}_{\rm{ef}})$.
The elements of the density matrix are then reconstructed with a maximum-likelihood method, assuming ideal tomography gates. 

To extend this QST procedure to two-qutrit density matrices, we perform two local tomography gates (from the 81 pairs of gates that can be formed from the single-qutrit QST gates) on transmon A and B, before extracting the state populations using the two-qutrit single shot measurement method described in Appendix~\ref{app:Readout}. 

To characterize the qubit state transfer from node A to node B we performed full quantum process tomography~\cite{Chuang1997}. We prepare each of the six mutually unbiased qubit basis states $\ket{g}$, $\ket{e}$, $(\ket{g}+\ket{e})/\sqrt{2}$, $(\ket{g}+i\ket{e})/\sqrt{2}$, $(\ket{g}-\ket{e})/\sqrt{2}$, $(\ket{g}-i\ket{e})/\sqrt{2}$ ~\cite{Enk2007}, transfer the state to node B, then independently measure the three-level density matrix at node A and node B with QST. We obtain the process matrix through linear inversion, from these density matrices.

%%%%%%%%%%%%%%%%%%%%%%%%%%%%%%%%%%%%%%%%%%%%%%%%%%%%%%%%%%%%%%%%%%%%%%%%%%%%%%%%%%%%%%%%%%%%%%%%%%%%%%%
\section{Two-Qutrit Entanglement}
\label{app:twoqutritEntanglement}
Due to a residual population of $3.5\%$ of the $\ket{f}$ level of the transmons after the entanglement protocol, the entangled state cannot be rigorously described by a two-qubit density matrix. To be concise we represent the reconstructed two-qutrit entangled state $\rho_{3 \otimes 3}$ (Fig.~\ref{fig:entanglementGEF}) by a two-qubit density matrix $\rho_{\rm{m}}$, that consists of the two-qubit elements of $\rho_{3 \otimes 3}$. This choice of reduction from a two-qutrit to a two-qubit density matrix conserves the state fidelity $\mathcal{F}^{s}_{\ket{\psi^+}}=\braket{\psi^+|\rho_{\rm{m}}|\psi^+}=\braket{\psi^+|\rho_{3 \otimes 3}|\psi^+}$, however, $\rho_{\rm{m}}$ has a non-unit trace. In addition, this reduction method gives a conservative estimate of the concurrence $\mathcal{C}(\rho_{\rm{m}})$, compared to a projection of $\rho_{3 \otimes 3}$ on the set of physical two-qubit density matrices.
To thoroughly verify the three-level bipartite entanglement, we use the computable cross norm or realignment (CCNR) criterion~\cite{Guhne2009}, which is well defined for multi-level mixed entangled states. The CCNR criterion states that a state $\rho$ must be entangled if $\rm{ccnr}=\sum_{\rm{k}}\lambda_{\rm{k}} > 1$ with $\rho=\sum_{\rm{k}}\lambda_{\rm{k}} \rm{G}_{\rm{k}}^{\rm{A}} \otimes \rm{G}_{\rm{k}}^{\rm{B}}$ and $\rm{G}_{\rm{k}}^{\rm{A(B)}}$ being an orthonormal basis of the observable spaces of $\mathcal{H}^{\rm{A}(\rm{B})}$. We obtain $\rm{ccnr}=1.612 \pm 0.003$ with the measured entangled state $\rho_{3\otimes 3}$, witnessing unambiguously the existence of entanglement of the prepared state.

%%%%%%%%%%%%%%%%%%%%%%%%%%%%%%%%%%%%%%%%%%%%%%%%%%%%%%%%%%%%%%%%%%%%%%%%%%%%%%%%%%%%%%%%%%%%%%%%%%%%%%%

\bibliography{Q:/USERS/kurphili/00Literature/RefDB/QudevRefDBRemoteEntanglementCopy}
\end{document}